\documentclass{moriond}

\def\be{\begin{equation}}
\def\ee{\end{equation}}
\def\bea{\begin{eqnarray}}
\def\eea{\end{eqnarray}}

\newcommand{\Photo}{}

\usepackage{amsmath}
\usepackage{amssymb}

\begin{document}
\vspace*{4cm}

\title{THE SUCCESS STORY OF SQUEEZED LIGHT}

\author{ ROMAN SCHNABEL }

\address{Institut f\"ur Laserphysik \& Zentrum f\"ur Optische Quantentechnologien, Universit\"at Hamburg, \\
Luruper Chaussee 149, 22761 Hamburg, Germany}

\maketitle
\abstracts{
Squeezed states of the optical field were theoretically described in the early 1970s and first observed in the mid 1980s. The measured photon number of a squeezed state is correlated with the measured photon numbers of all other squeezed states of the same ensemble, providing sub-Poissonian statistics. Today all gravitational-wave observatories use squeezed light as the cost-efficient alternative to further scaling up the light power. This user application of quantum correlations was made possible through dedicated research and development of squeezed light between 2002 and 2010.
}

\section{Introduction} \vspace{-1mm}
A mode of the optical field is in a coherent state $| \alpha \rangle$ (Glauber state)\;\cite{Glauber1963}, if photon number measurements on many identical such modes provide (i) an average photon number $\overline{\!N} = |\alpha|^2$ and (ii) a Poissonian counting statistics indicating random, mutually independent, uncorrelated photon events. In the limit of large displacements ($|\alpha| \gg 1$), the Poissonian distribution is approximately given by a Gaussian with a standard deviation of $\sqrt{\overline{\!N}}$. A largely displaced coherent state is experimentally certified if the measured noise power, which is proportional to the variance of the quantum uncertainty, is proportional to the optical power.

The description of a squeezed state $| \alpha, r, \theta \rangle$ requires two more parameters. $r > 0$ is the squeeze parameter, and $\theta$ is the squeeze angle\;\cite{Stoler1970}. The noise power of a squeezed state is reduced by the factor ${\rm exp}(2r)$ in comparison to a coherent state of the same displacement. $\theta = 0$ is assigned to an amplitude quadrature squeezed state; $\theta = \pi/2$ is assigned to a phase quadrature squeezed state~\cite{Walls1983,Breitenbach1997}. 
Squeezed states of the optical field were first theoretically described by D.~Stoler and E.\;Y.\;C.~Lu in the early 1970s\;\cite{Stoler1970,Lu1971} and a few years later by H.\;P.~Yuen\;\cite{Yuen1976}. 

In 1981, C.\;M.~Caves proposed squeezed states for gravitational-wave (GW) detection to work around high-power lasers~\cite{Caves1981}, which were not available in the 1980s. The first measurement of squeezed states was achieved in the mid 1980s~\cite{Slusher1985}, with an inferred squeeze factor of up to 3\,dB ($10\cdot{\rm log}_{10}\{\,{\rm exp}(2r)\!\approx\!0.5\,\}$)~\cite{Wu1986}. 3\,dB of amplitude squeezing improves the signal to shot-noise ratio as much as doubling the coherent light power. In subsequent years, increasing the squeeze factor turned out to be a difficult task. By the end of the century, up to about 6\,dB was observed in a few research labs~\cite{Polzik1992,Breitenbach1997b,Lam1999}. 
Much more progress was achieved on ultra-stable single mode lasers~\cite{Freitag1995}. On top of that, the concept of enhancement cavities was invented\;\cite{Drever1983,Fritschel1992,Mizuno1993} promising hundreds of kilowatts in the arms of GW detectors. 
Furthermore, all squeezing experiments were performed at MHz sideband frequencies, whereas GW detection requires squeezing in the audio-band. The application of squeezed light in GW observatories seemed more an academic exercise than a useful task.
However, in the early 2000's it became clear that further scaling up the light power in GW observatories beyond 100\,kW causes severe technical problems\;\cite{Strain1994,Braginsky2001}. Dedicated research and development of squeezed light for GW observatories then started.

\section{Measuring squeezed light} 
\begin{figure}[h!!!!!!!!!!!!!!!!!!!!!!!!!!!!!!]
\vspace{3mm}
\centerline{\includegraphics[width=0.8\linewidth]{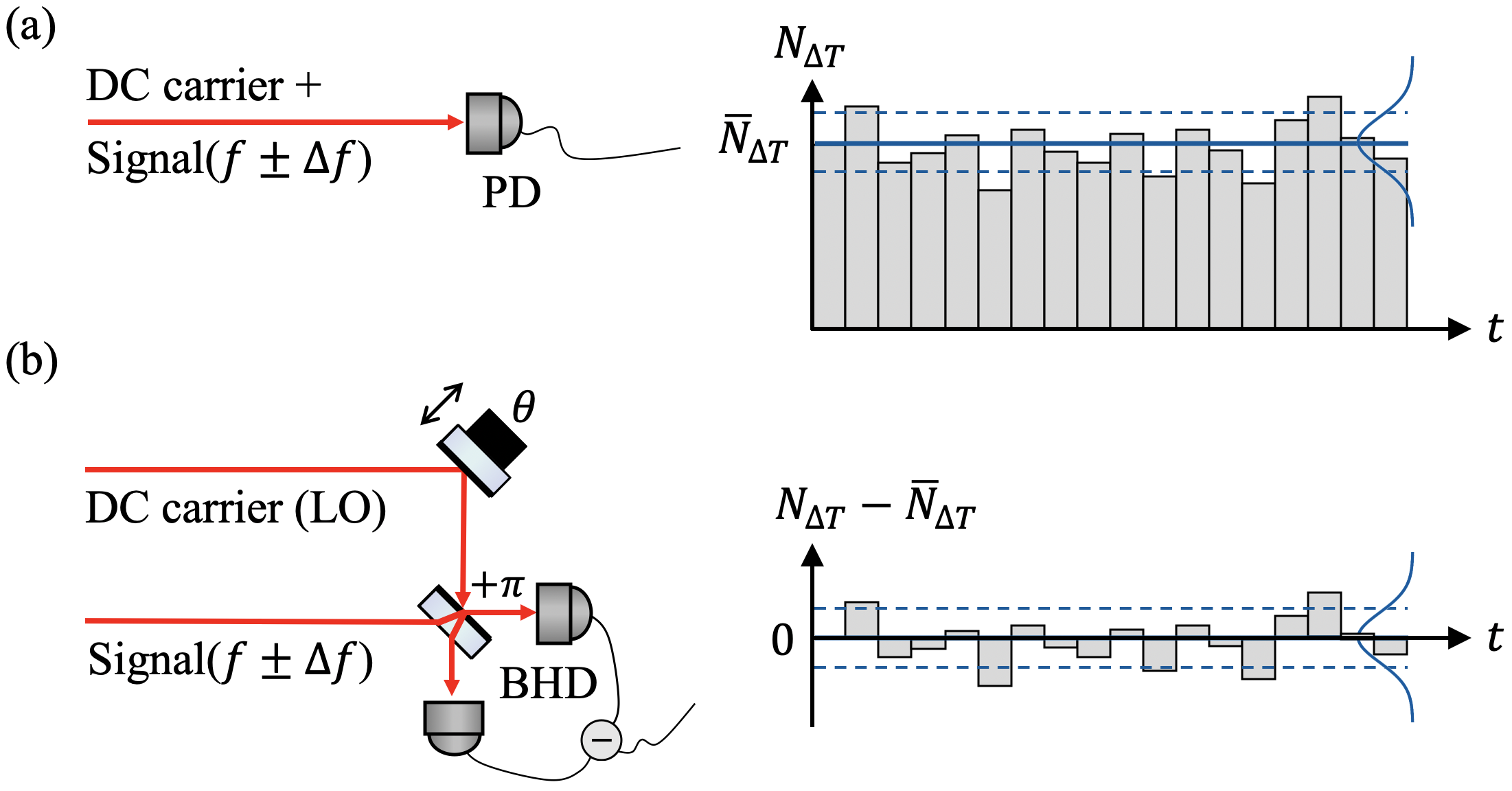}}
\caption{
Photo diode (PD) arrangements for measuring optical signals with squeezed quantum noise carried by quasi-monochromatic (DC) light.
(a), left: Signals are modulations in some expected frequency band $f \pm \Delta f$.  If the signals are amplitude (power) modulations and the quantum noise amplitude squeezed (with respect to the carrier field) a single photo diode serves as a measurement device. This arrangement cannot measure phase quadrature modulations/squeezing. 
Right: Noisy photon-number time series without any obvious signal.
(b), left: The output of a balanced homodyne detector (BHD) is the difference of the voltages of two identical PDs. A BHD is able to measure arbitrarily weak signal fields, including ground state uncertainties (`no' signal beam) and squeezed uncertainties. The required optical amplification is provided by the spatially overlapped local oscillator (LO) beam of the same wavelength with a power of the order of 10\,mW. The power of the signal beam should be smaller than 1\% of the LO power. In contrast to (a), a BHD can deliberately set the differential path length of signal and LO field. This feature allows for measuring the phase quadrature.
Right: Same photon-number noise as in (a) but with the average value $\overline{\!N}_{\!\Delta\!\,T}$ removed. Note that the optically amplified signal in the two BHD arms adds up due to the phase flip ($+\pi$) for one of the beam splitter reflections. The phase flip is a necessary consequence of energy conservation.
}
\label{fig:1}
\end{figure}

\vspace{-1mm}
Squeezed light is measured with a conventional photo diode (PD) or with two such PDs in a balanced arrangement, see Fig.\,\ref{fig:1}. PDs are usually operated with incident optical powers of the order of 10\,mW, where their dark noise can be mostly neglected. PDs do not have single photon resolution, but provide a unity-gain (one-to-one) mapping of a large-number photon statistics onto an identical photo-electron statistics. Quantum efficiencies of greater than 99\% have been realised\;\cite{Vahlbruch2016}.
The photoelectric output voltage of a PD is proportional to the absorbed photon number per measuring time interval $\Delta T$. For continuous-wave light, $\Delta T$ is usually given by the response time of the PD and its electronic amplification stage. 
It is important to note that after a time-frequency analysis, the photoelectric voltage is proportional to the product of the amplitude of the DC field $\alpha_0$ (carrier) and the signal field's amplitude $X_{f,\Delta f} (t)$ being the (time-dependent) depth of the beam's amplitude modulation at sideband frequency $f$ over the resolution bandwidth $\Delta f$~\cite{Schnabel2017}. For a signal analysis in time-frequency representation, see for instance the bottom row of Fig.\,1 of Ref.\;\cite{GW150914}. Note that the signals may be time dependent but the `squeezing' is usually stationary. More information about the measurements of signals with squeezed quantum noise can be found in Ref.\;\cite{Schnabel2017}.

The balanced homodyne detector (BHD) is the superior detection arrangement if the signal beam power is significantly below the power a photo diode can handle and if a coherent local oscillator (LO) beam of higher power is available to be spatially overlapped with the signal field on a balanced beam splitter. A BHD can measure the electric field of a ground state (vacuum state), as well as that of squeezed vacuum states. It can measure the field at any quadrature angle, from the amplitude $X_{f,\Delta f} (t)$ to the phase quadrature $Y_{f,\Delta f} (t)$. Technical noise on the LO beam cancels out due to the balanced arrangement.

\section{Optical modes in squeezed states} 
\begin{figure}[h!!!!!!!!!!!!!!!!!!!!!!!!!!!!!!]
\centerline{\includegraphics[width=0.85\linewidth]{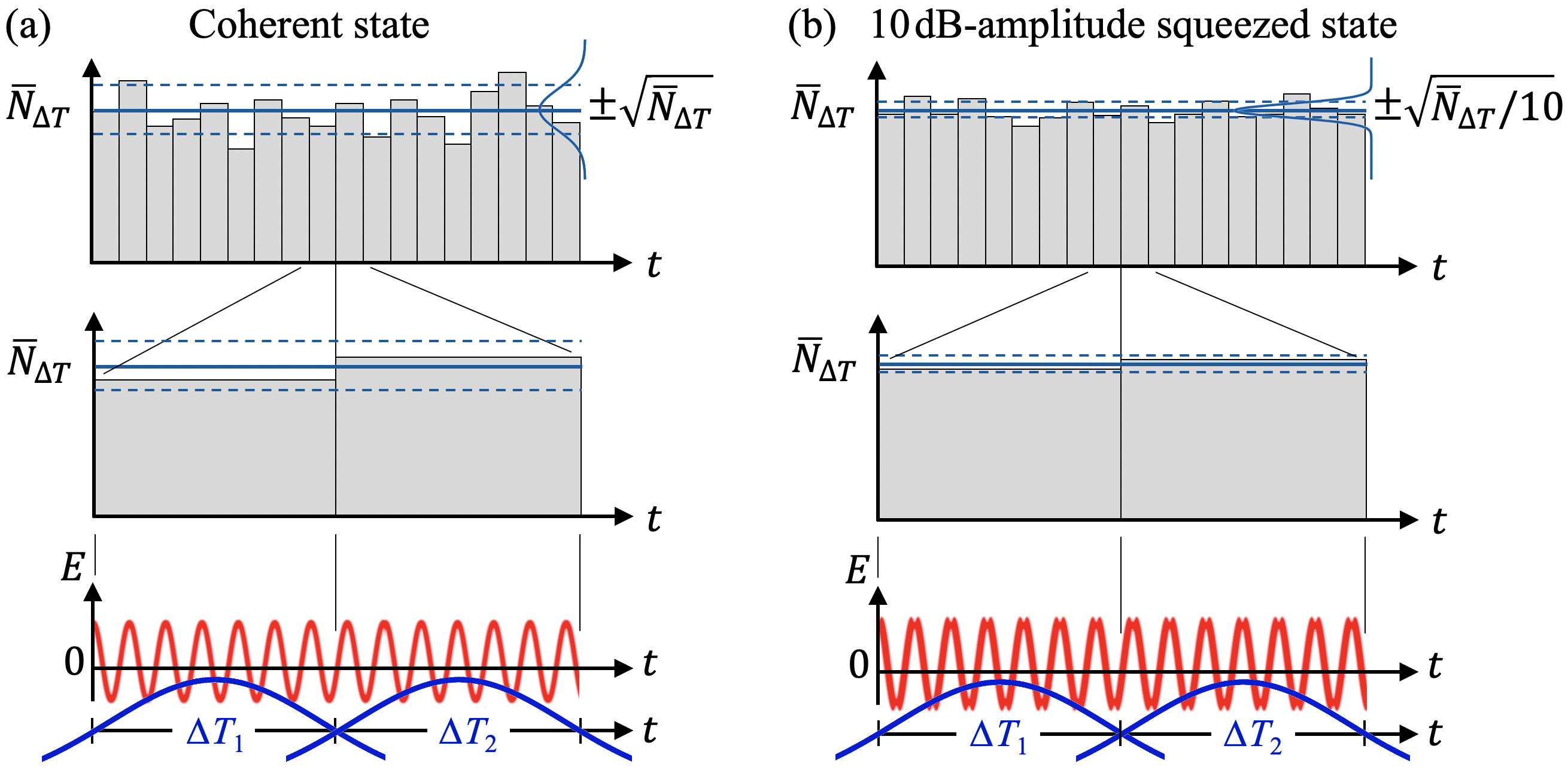}}
\caption{
Top row: measured photon numbers of short Gaussian measurement modes (bottom row). 
(a) illustrates an ensemble of such modes in a coherent state. (b) illustrates the same but for a 10\,dB amplitude squeezed state. The photon statistics in the top right panel is sub-Poissonian. This feature has been used for improving the signal to noise ratio in GW observations since 2019.
Bottom row: Shown are the monochromatic fields including their quantum uncertainties. (a) Phase independent quantum uncertainty of a displaced coherent state. (b) Same light field but with amplitude squeezed quantum uncertainty. The Gaussians illustrate the sequence of measurement modes that are `cut out' by the measuring apparatus, and to which a photo voltage value gets assigned.
}
\label{fig:2}
\end{figure}

\vspace{-1mm}
Fig.\,\ref{fig:2} illustrates the definition of the modes of the light that are either in a coherent state (a), in a squeezed state (b), or in any other state. 
Firstly, photon numbers are subsequently measured in short time intervals $\Delta T$ (first row). Every time interval has a temporal mode function that is defined by the measurement device (and not by the continuous-wave light). The mode function may have a Gaussian profile (bottom row). 
The coherent or squeezed states we are talking about are the quantum states of these short modes. Together they define an ensemble of identical modes in identical states. Note that the mode of the light before the measurement is a different one. It has a temporal mode function of half width $\Delta t$ that is much larger than $\Delta T$. $\Delta t$ is given by the Fourier transform of the quasi-monochromatic spectrum of the light\;\cite{Schnabel2020}.

Fig.\,\ref{fig:2} thus illustrates an example of the quantum measurement process. Since the initial mode of the light is quasi-monochromatic, it has a large  coherence time $\Delta t$; the measurement device cuts out many short modes of the same length; the reduced length enforces the short modes having an increased spectral width, which is described by the Fourier transform limit. Since the initial mode has a narrow band spectrum, the short modes are Fourier transform limited. Furthermore, they are all indistinguishable; they are measured at different times, but this fact does not make them distinguishable because the different times are all within the coherence time of the initial mode. The measurement thus defines an ensemble of identical Fourier-limited modes of length $\Delta T$ that are in identical states.

\vspace{3mm}

\vspace{-1mm}
\section{``Weird'' quantum correlations}
\vspace{-1mm}
The coherent-state quantum statistics in Fig.\,\ref{fig:2}\,(a) can be well described by the semi-classical model as follows.
(i) The initial continuous-wave light (bottom row, left) must have a long coherence time $\Delta t \gg \Delta T$ because it is quasi-monochromatic. 
(ii) According to the mathematics of the Fourier transform, any energy of the mode cannot be localised more precisely than $\Delta t$. 
(iii) From this follows that a photon event within some measurement mode $\Delta T_j$ must appear in a truly random fashion. This model explains the occurrence of the Poissonian distribution, which is the counting statistics of mutually independent uncorrelated events.\\
The fact that measurements on quasi-monochromatic light can produce photon numbers in short time windows with a sub-Poissonian statistics, however, is weird. The mathematics of the Fourier transform again requests that the energy of the beam is smoothly distributed over $\Delta t$,  which guarantees a truly random occurrence of photons within $\Delta T \ll \Delta t$.
But a sub-Poissonian statistic only seems possible if the photons are `deliberately' fairly evenly distributed among the modes of length $\Delta T_i$, which seems to require an unknown force. The phrase ``squeezed states cannot be described by a semiclassical model'' refers to this weirdness. 
Squeezed states have `quantum correlations' and belong to the class of `nonclassical' states. 
A more detailed description of my argumentation is given in Ref.\;\cite{Schnabel2020}

\section{Research and development between 2002 and 2010} \vspace{-1mm}
Back in 2002, it was unclear whether quantum noise squeezing was feasible not only at MHz sideband frequencies but also in the audio-band, as required for GW observatories. Progress in answering this question prompted the new question of whether a robust servo-control technique is possible to phase-control squeezed vacuum states that are not co-propagating with DC carrier light.
It was also unclear whether applying squeezed states are beneficial to interferometers that used enhancement cavities. 
Issues with the signal recycling cavity\;\cite{Meers1988} were discussed in Ref.\;\cite{Chickarmane1998}.  
Last but not least was it unclear whether relevant squeeze factors of the order of `ten' and above could be achieved at all.

\begin{figure}[h!!!!!!!!!!!!!!!!!!!!!!!!!!!!!!]
\vspace{3mm}
\centerline{\includegraphics[width=1\linewidth]{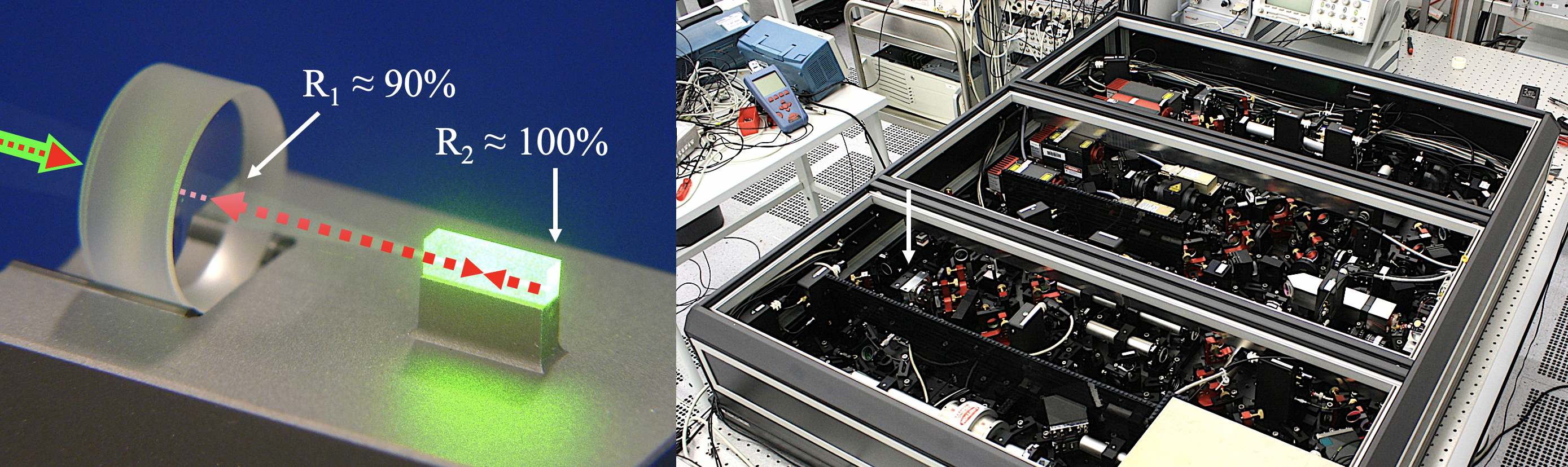}}
\vspace{0mm}
\caption{Left: Standing-wave squeezing resonator constructed from a coupling mirror with reflectivity $R_1$ at the optical frequency of the squeezed light $f$ and the back of a non-linear crystal with an almost perfect reflectivity $R_2$. The cavity resonance at $f$  (dashed line) is in the vacuum state when the pump at $2f$ is off (green solid arrow). If $f$ encounters a cavity resonance and the pump field is switched on, a squeezed vacuum field leaves the resonator to the left. The two optical components shown are clamped to a rigid mount. The resonator length is servo controlled, e.g.~by transmitting a weak phase-modulated coherent beam through the resonator (not shown). The crystal's temperature is stabilised to its phase matching temperature. Right: First squeeze laser for 24/7 operation built by H. Vahlbruch and other members of my group. The squeeze laser has been being used as part of GEO\,600 since 2010. The arrow points to the squeezing resonator. It has provided a squeeze factor of up to 10\,dB over the audio band. 
}
\label{fig:3}
\end{figure}

By far the most efficient process for squeezed light production is resonator-enhanced, degenerate type\,I parametric down-conversion in nonlinear crystals\;\cite{Wu1986}. An example resonator is shown in Fig.\,\ref{fig:3}, left. In contrast to an optical-parametric oscillator, the device is operated slightly below lasing threshold. The process can be well described physically using the methods of nonlinear cavity optics: 
A quasi-monochromatic beam of light at optical frequency $2f$ modulates the dielectric polarisability of the medium (e.g.~MgO:LiNbO$_3$ or KTiOPO$_4$) which results in a $2f$ modulation of the magnitude of the quantum uncertainty. Phases of damping are squeezed; phases of amplification anti-squeezed, see Fig.\,14 of Ref.\;\cite{Schnabel2017}. The crystal is located inside a single-ended optical resonator. Impedance-matching is achieved for the damped field quadrature and simultaneously lasing is achieved for the amplified quadrature, if the field attenuation due to the parametric round-trip damping matches the transmission of the resonator's coupling mirror, see Fig.\,19 of Ref.\;\cite{Schnabel2017}. For an optimal squeezing resonator, the pump intensity is adjusted slightly below that point. For further information, I refer to the review\;\cite{Schnabel2010}.

\begin{figure}[h!!!!!!!!!!!!!!!!!!!!!!!!!!!!!!]
\vspace{3mm}
\centerline{\includegraphics[width=0.9\linewidth]{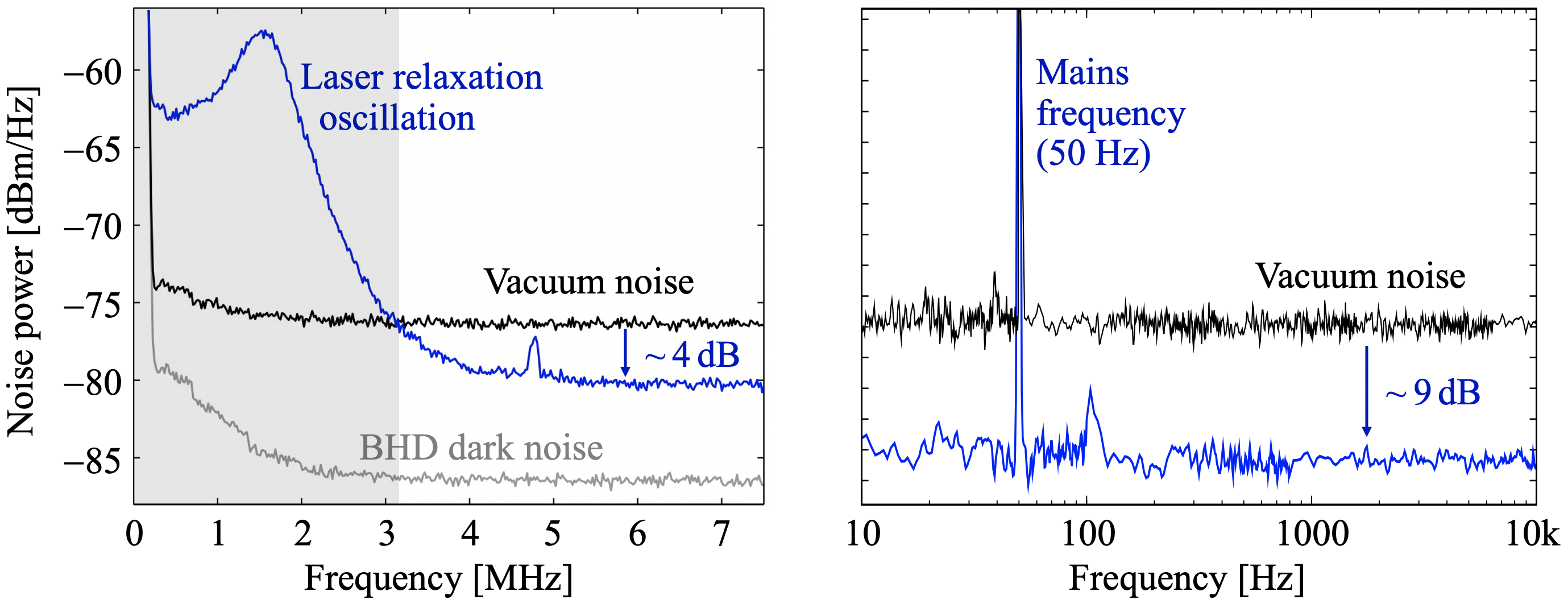}}
\vspace{-2mm}
\caption{Measured noise power spectral densities of continuous-wave carrier light at 1064\,nm. 
Left: Typical measurement around 2002. Squeezed states were observed at frequencies above a few MHz. At lower frequencies, laser noise from the co-propagating control field as well as back-scattered light buried the squeezing. 
Right: Measurement of squeezed states over the entire signal spectrum of ground based GW observatories from 10\,Hz to 10\,kHz reported in [Vahlbruch2010]. This result was achieved in 2009 by: avoiding DC carrier light on the squeezed field, using a 40\,MHz shifted single-sideband field that was transmitted through the squeezing resonator for creating a phase control error signal, the availability of photo diodes with 99.5\% quantum efficiency at 1064\,nm, using periodically poled KTP crystals with low absorption, achieving an interference contrast at the balanced homodyne detector of $>99$\%, using super-polished mirrors and lenses, and shielding the photo diodes against back-scattered light at 1064\,nm.
}
\label{fig:4}
\end{figure}
\subsection{Squeezed quantum noise at audio-band frequencies} \vspace{-1mm}
In 2001, I was a postdoctoral member of a team in the group of H.-A. Bachor and P.\;K. Lam at the Australian National University (ANU) that aimed for the experimental demonstration of quantum teleportation. When W.\,P. Bowen and myself combined the beams from two squeezing resonators on a balanced beam splitter with zero differential phase, we found in one of the beam splitter outputs the frequencies of squeezing expanded below the MHz range\;\cite{Bowen2002} -- in contrast to Fig.\,\ref{fig:4}\,(a). The other beam splitter output showed deteriorated squeezing below 3\,MHz. The reason was the destructive interference of the co-propagating weak DC control fields in the output port of the improved spectrum. The destructive interference included laser noise that came with the control fields\;\cite{Schnabel2004}. We concluded that the required error signals for locking the length of the squeezing resonator and for setting the phase of the BHD had to be generated without weak DC control fields entering the BHD through its signal port. Later, another team at the ANU observed for the first time squeezed states at audio-band frequencies down to 280\,Hz\;\cite{McKenzie2004} with a slowly drifting length of the squeezing resonator and with the LO phase $\theta$ -- see Fig.\,\ref{fig:1} -- locked on lowest quantum noise variance (`quantum noise locking')\;\cite{McKenzie2005}.
My team at Leibniz Universit\"at Hannover invented an alternative locking scheme based on a classical single sideband modulation. The sideband was produced by an acousto-optical modulator at 40\,MHz and transmitted through the squeezing resonator. The parametric down-conversion process turned the single sideband into an unbalanced pair of sidebands at $\pm 40$\,MHz, from which an error signal for locking the BHD could be created. This `coherent locking'\;\cite{Vahlbruch2006,Chelkowski2007} is as stable as other radio-frequency phase control schemes and allowed for the first time the generation of a broad-band squeezing spectrum that covered the entire audio band from 10\,Hz to 10\,kHz\;\cite{Vahlbruch2007}. In 2010, coherent locking was used to realise the squeezing enhancement of GEO\,600\;\cite{Vahlbruch2010,LSC2011}, which was operated as a GW observatory at that time\;\cite{Grote2013}. 
Today, coherent locking is used in all squeezing enhanced GW observatories to lock the squeezed quadrature to the quadrature of the GW signal.

\subsection{Compatibility of squeezed light with interferometers being enhanced by cavities} \vspace{-1mm}
Fig.\,5 shows the optical topology of a cavity-enhanced Michelson laser interferometer for the observation of gravitational waves, as realised in the current Advanced LIGO and Advanced Virgo GW observatories. While the arm resonators, power-recycling as well as signal-recycling were in the original designs, squeezed light injection was not planned initially, since the squeezed light technology was not mature at that time. Squeezed light injection was included after several proof-of-principle experiments and the highly successful implementation in GEO\,600 in 2010.

A Michelson interferometer achieves its highest signal to photon-shot-noise ratio when operated close to a dark signal output port, see for instance Eq.\,(5.41) in Ref.\;\cite{BachorRalph2019}. In this setting the squeezed light must be injected into the signal port\;\cite{Caves1981}, as shown in Fig.\,5. 
The compatibility of squeezed light injection with power-recycling was experimentally proven in 2002 in Ref.\;\cite{McKenzie2002}. The power-recycling mirror was found to be able to reduce the optical loss on the squeezed state.
The compatibility of squeezed light injection with signal-recycling was theoretically investigated in detail by J. Harms {\it et al.}\;\cite{Harms2003}, which included the optical spring effect due to radiation pressure forces when the signal-recycling cavity is detuned from the carrier and tuned to a frequency in the lower signal sideband (`red-detuned'). Subsequently, an experiment demonstrated compatibility of signal recycling. The signal-recycling cavity was operated detuned and the compensation of the unwanted frequency-dependent phase-space rotation of the squeezed light demonstrated\;\cite{Vahlbruch2005}. 
A test in a prototype GW detector with suspended optics was achieved by a team at the LIGO laboratories\;\cite{Goda2008,Schnabel2008}. 
\begin{figure} 
	\begin{minipage}{0.55\textwidth}
		\caption{Optical topology of current GW observatories. The purpose of the arm resonators is to increase the light power and to increase the interaction time of the light with the gravitational wave. The power-recycling mirror creates another resonator together with the highly reflective interferometer being operated close to a dark output port. It further builds up the light power in the arms without affecting the GW signal, since the latter is a differential arm length signal and as such fully couples to the output port. The signal-recycling mirror together with the interferometer forms a third type of cavity with the purpose to optimise the effective signal bandwidth of the arm resonators. In LIGO and Virgo, it effectively broadens the bandwidth of the arm resonators (`resonant sideband extraction'). In order to squeeze the quantum noise of the (differential arm length) GW signal, the squeezed light needs to be injected into the signal port. It should be retro-reflected towards the photo diode with low loss. Red lines: Continuous-wave light at 1064\,nm. PBS: polarising beam splitter.}
	\end{minipage}\hfill
	\begin{minipage}{0.4\textwidth}
		\includegraphics[width=1\linewidth]{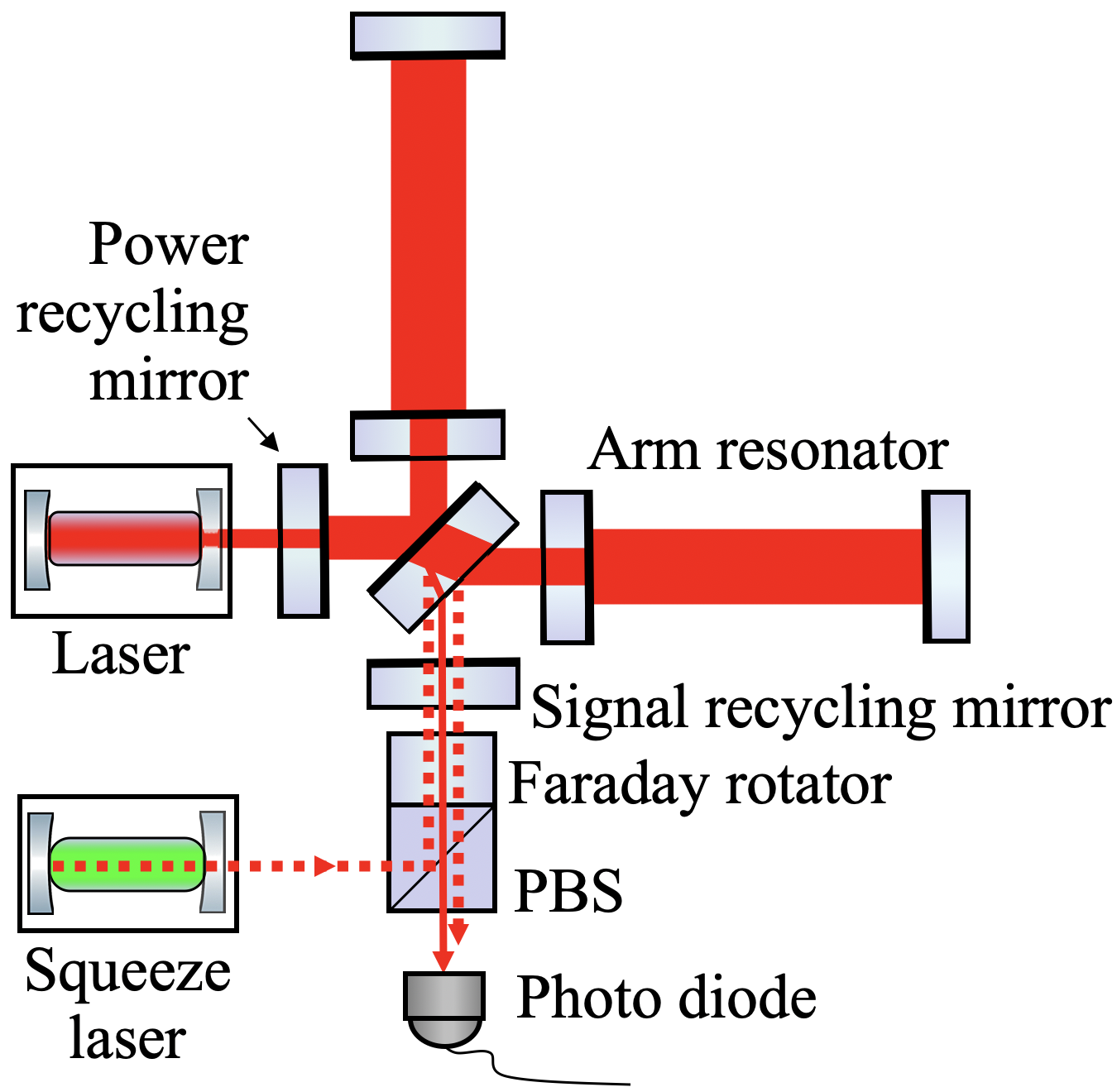}
	\end{minipage}
\end{figure}

\subsection{Squeeze factors $\,\ge 10\,{\rm dB}$} \vspace{-1mm}
In 2007, a team in the group of A. Furusawa at the University of Tokyo reported on the observation of 9\,dB of squeezing on continuous-wave light at 860\,nm\;\cite{Takeno2007}. One year later my group was able to report 10\,dB at 1064\,nm\;\cite{Vahlbruch2008,Polzik2008}. The improvements compared to previous experiments were possible due to higher phase stabilities, mirrors and lenses with super-polished surfaces, and photo diodes of improved quantum efficiencies. 
Our squeezing value at 1064\,nm was mainly limited by the nevertheless imperfect quantum efficiency of $95 \pm 2 \%$. In 2010, we could improve the squeeze value to 11.5\,dB\;\cite{Mehmet2010} by using photo diodes whose quantum efficiency we estimated to $98 \pm 2 \%$ at that time. These photo diodes were fabricated for us by the Fraunhofer-Heinrich-Hertz-Institut in Berlin from a custom-designed InGaAs-wafer. The performance of these photo diodes was so good that GEO\,600, LIGO and Virgo also started to use them. In a later experiment we were able to measure their quantum efficiency more precisely to $99.5 \pm 0.5 \%$\;\cite{Vahlbruch2016}.
\begin{figure} \vspace{-2mm}
	\begin{minipage}{0.51\textwidth}
		\caption{Shown are the results from the first observation of 10\,dB of quantum noise squeezing [Vahlbruch2008]. Increasing the optical loss on the squeezed light reduced the squeeze and anti-squeeze factors in a characteristic way, which precisely corresponded to the theoretical model assuming a parametric gain of 63. The small rectangles corresponded to the measurement data and their sizes to the error bars. Assuming that optical loss was the only source of decoherence allowed us to quantify the intrinsic loss in our setup to 8.6\%. Also compatible with our measurement data was an intrinsic optical loss of at least 5.6\% plus phase noise of up to 1.2$^\circ$ at most. The squeeze factors were measured at sideband frequencies around 5\,MHz.}
	\end{minipage}\hfill
	\begin{minipage}{0.44\textwidth}\vspace{-2mm}
		\includegraphics[width=1\linewidth]{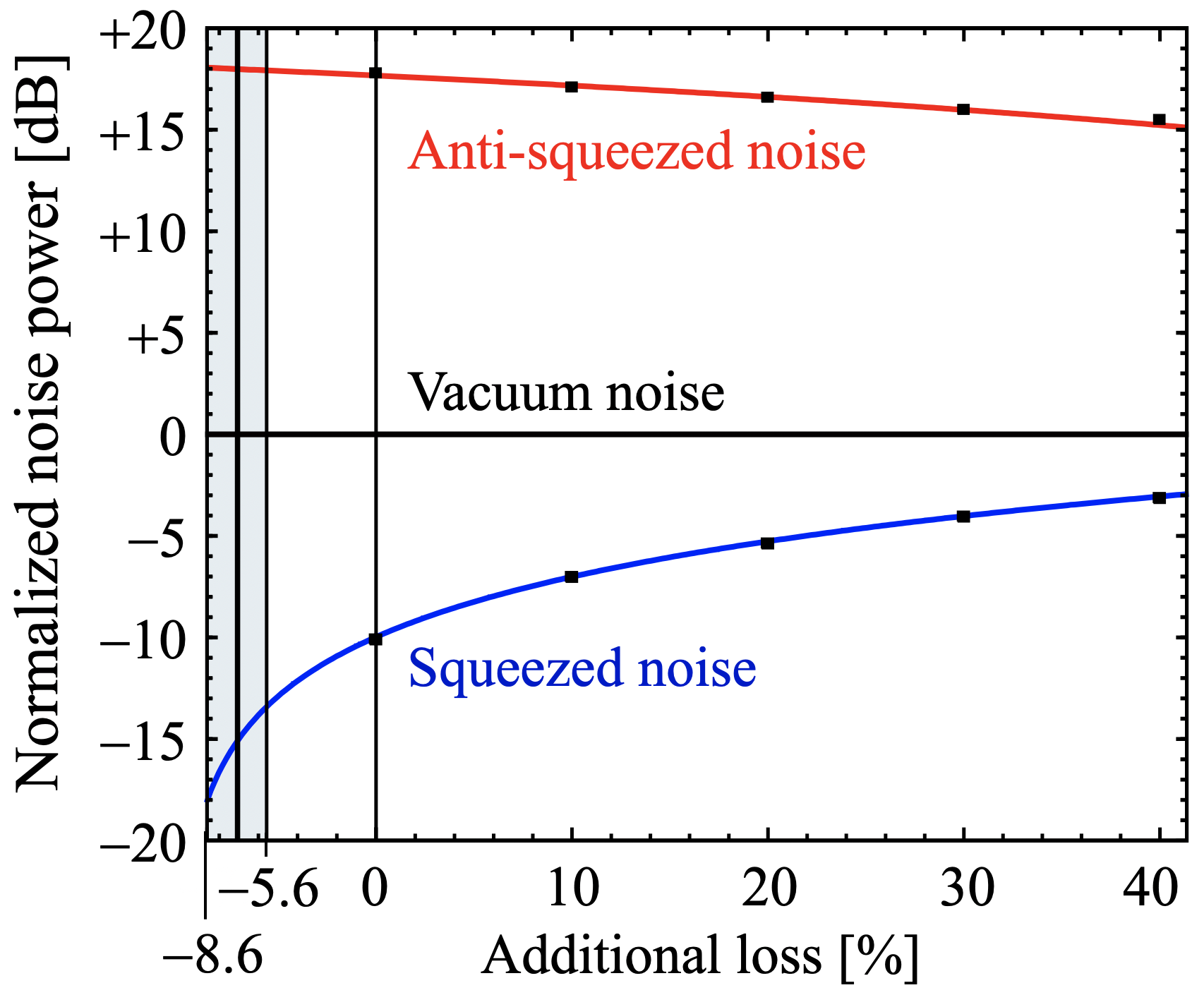}
	\end{minipage}
\end{figure}

\vspace{0mm}
\section{The `squeeze laser' for improving GW observatories} \vspace{-1mm}
The photograph in Fig.\,\ref{fig:3} (right) shows the first squeeze laser\;\cite{Vahlbruch2010} that was integrated in a GW observatory\;\cite{LSC2011}. The installation on site happened in 2010, when GEO\,600 was operated as a GW observatory\;\cite{Grote2013} --- Today, GEO\,600 is basically used for prototyping new technologies. The squeeze laser has been providing up to 10\,dB of squeezing over the audio band. Fig.\,\ref{fig:4} (right) shows an example spectrum showing up to 9\,dB, which was limited by the not-optimal BHD used. Optical losses inside GEO\,600 initially reduced the effective improvement on site to 3.5\,dB\;\cite{LSC2011}. Recently, up to 6\,dB of nonclassical improvement of GEO\,600's signal to noise ratio with the same squeeze laser was reported\;\cite{Lough2021}.

The success of the squeeze laser in GEO\,600 changed the initial plans for Advanced LIGO and Advanced Virgo. Around the end of the last century they were designed without squeezed light injection, because the squeeze technology was not mature and further scaling up the light power in the arms of GW observatories seemed possible. Since 2019, Advanced LIGO and Advanced Virgo use squeezed light during all GW observation runs\;\cite{Abbott2021o31}. While the light powers in the arms could not yet be increased to the design values, squeezed light helps for reaching the design sensitivities.

\section{Squeezed light for future GW observatories -- outlook} \vspace{-1mm}

\emph{Squeezed fields with frequency dependent squeeze angle} --
In the near future, quantum radiation pressure noise (RPN) that acts on the four test masses in the GW observatory arms needs also to be squeezed to increase the sensitivity at signal frequencies below 50\,Hz. The simultaneous squeezing of photon shot noise at the photo diode and RPN at the mirrors is possible by exploiting quantum correlations\;\cite{Jaekel1990} and can be realised by 300\,m long high-quality filter cavities\;\cite{Kimble2001,Barsotti2019}. The installations of these filter cavities are immediate projects at the current GW observatories\;\cite{Zhao2020,McCuller2020}.
\\[1mm]
\emph{Short wavelength infrared squeezed light} -- 
The next generation of GW observatories might be operated at slightly longer infrared wavelengths, e.g.~at 1550\,nm or 2\,$\mu$m. This move will enable mirror materials and mirror coatings with lower levels of thermal noise as determined for instance by Brownian vibrations of the mirror surfaces. At 1550\,nm, a squeeze factor of up to 13\,dB was realised\;\cite{Mehmet2011,Schoenbeck2018}. At 2\,$\mu$m and slightly above the values are significantly lower\;\cite{Mansell2018,Darsow-Fromm2021}, limited by the quantum efficiencies of the photo diodes currently available.
\\[2mm]
\emph{Squeeze operation inside GW observatories} --
GW observatories beyond the next generation could benefit from an additional squeeze crystal located inside the GW interferometer\;\cite{Rehbein2005,Peano2015}. Theoretical models as well as early experiments predict the possibility of shaping\;\cite{Somiya2016,Adya2020} and widening the observatory's signal response\;\cite{Korobko2017,Korobko2019}. A nonlinear crystal inside the signal-recycling cavity will certainly increase the optical loss due to crystal absorption and due to imperfect anti-reflection coatings on the crystal, however, it seems realistic with dedicated research and development.\\

\vspace{0mm}
\section*{Acknowledgments}
\vspace{-2mm}
The work by R.S. from 2003 to 2009 described here was supported by the Deutsche Forschungsgemeinschaft (DFG, The German Research Foundation)  
through the collaborative research centre SFB\,407.

\end{document}